\documentclass[acus]{JAC2000}

\def\bold#1{\setbox0=\hbox{$#1$}%
     \kern-.025em\copy0\kern-\wd0
     \kern.05em\copy0\kern-\wd0
     \kern-.025em\raise.0333em\box0 }

\catcode`\@=11 
\def\lsim{\mathrel{\mathpalette\@versim<}}
\def\gsim{\mathrel{\mathpalette\@versim>}}
\def\@versim#1#2{\vcenter{\offinterlineskip
        \ialign{$\m@th#1\hfil##\hfil$\crcr#2\crcr\sim\crcr } }}
\catcode`\@=12 

\usepackage{graphicx}

\begin{document}
\title{\flushright{W10}\\[15pt] \centering 
A POSSIBLE RESOLUTION OF THE 
$\,\bold{e^+e^- \rightarrow \bar N N}\,$ PUZZLE\thanks{Invited 
talk at the workshop \ {\em $``e^+ e^-$ Physics at Intermediate Energies"},
SLAC, April 30 - May 2, 2001.}}

\author{Marek Karliner\thanks{\tt e-mail: marek@proton.tau.ac.il}, 
School of Physics and Astronomy,\\
Raymond and Beverly Sackler Faculty of Exact Sciences\\
Tel-Aviv, Israel}

\maketitle

\begin{abstract}
We sketch some recent ideas proposed as the \hbox{mechanism} behind
the puzzling experimental
results on baryon-antibaryon production in $e^+e^-$ annihilation
close to threshold.
The essential new point in the proposed \hbox{mechanism} is that 
it is a two-stage process, with a coherent state of pions serving as an
intermediary between $e^+e^-$ and the baryon-antibaryon system.
Skyrmion-antiskyrmion annihilation is proposed as a concrete computational
framework for a quantitative description of the baryon-antibaryon
 annihilation.
We also point out the possible connection to similarly puzzling data on
baryon-antibaryon production in photon-photon collision.
\end{abstract}

\section{INTRODUCTION -- THE PUZZLE}

The FENICE data \cite{Antonelli:1998fv}, 
on the reaction 
\hbox{$e^+e^- \rightarrow \bar n n$} close to threshold,
together with earlier analogous measurements for the proton
\cite{Antonelli:1994kq}-\cite{Bisello:1990rf}
indicate that at threshold
\hbox{$\sigma(e^+e^- \,\rightarrow\, \bar p p)/
\sigma(e^+e^- \,\rightarrow\, \bar n n) \approx 1$}.
In other words, the timelike form factors of the neutron and the proton
are approximately equal at $q^2\gsim M^2_N$.

\begin{figure}[thb]
\centering
\includegraphics*[width=85mm]{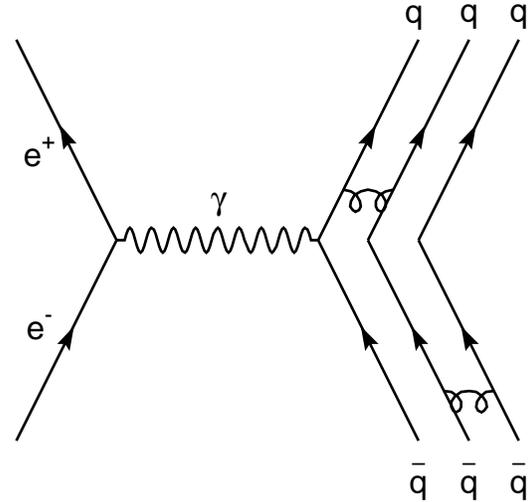}
\caption{Feynman diagram corresponding to the naive perturbative 
description for 
$e^+e^- \,\rightarrow\, \bar N N$.}
\label{ee-fig}
\end{figure}

This is a very surprizing and puzzling result 
and it is hard to understand in
the conventional perturbative
picture of baryon-antibaryon production in 
$e^+e^-$ annihilation, as shown in Fig.~\ref{ee-fig}.

\subsection{The perturbative picture}
In the naive perturbative description of the $e^+e^-$ annihilation into
baryons, the virtual timelike photon first makes a 
``primary"
$\bar q q$ pair, which
then ``dresses up" with two additional quark-antiquark pairs which pop up
from the vacuum. The ``dressing up" is a QCD process, which does not
distinguish between $u$ and $d$ quarks, since gluon couplings are
flavor-blind. Thus in the conventional picture the only difference between
proton and neutron is through the different electric charge of the primary 
$\bar q q$ pair. The total perturbative cross-section 
$\sigma_{\!\scriptscriptstyle PT}$
at a given {\it CM} energy
is obtained by superposing the
amplitudes with different flavors $q$ in the primary 
$\bar q q$ pair and squaring the result,
\begin{equation}
\sigma_{\!\scriptscriptstyle PT}(e^+e^- \,\rightarrow\, \bar N N) 
\propto \displaystyle \left\vert \sum_{q\in N} Q_q a^N_q(s)
\right\vert^2
\label{QED-NN}
\end{equation}
where $a^N_q(s)$ denotes the amplitude at $E^2_{CM}=s$ 
for making the baryon $N$ with a given
the primary flavor $q$.
These amplitudes are determined by
the baryon wavefunctions.

Since the wave functions of the baryon octet have a mixed symmetry,
the amplitudes $a^N_q(s)$ tend to be highly asymmetric. Thus for example
in the Chernyak-Zhitnitsky proton wave function 
\cite{Chernyak:1984ej}
the $u$ quark
dominates, i.e.  $a^p_u \approx 1$, $a^p_d \ll 1$ and similarly
$a^n_d \approx 1$, $a^n_u \ll 1$. In such a limiting case we have
\begin{equation}
{\sigma_{\!\scriptscriptstyle PT}(e^+e^- \,\rightarrow\, \bar p p)
\over
\sigma_{\!\scriptscriptstyle PT}(e^+e^- \,\rightarrow\, \bar n n) }
\quad \longrightarrow \quad
{Q_u^2\over Q_d^2} = 4
\label{PT-NN}
\end{equation}
While this is an extreme case, on general grounds
we expect that $u$ dominates in the proton and
$d$ in the neutron, so 
$\sigma_{\!\scriptscriptstyle PT}(\bar p p)/
 \sigma_{\!\scriptscriptstyle PT}(\bar n n)\gg 1$,
Intuitively one can understand this perturbative
result directly from Fig.~\ref{ee-fig},
by recalling that the average charge squared of quarks in the proton
is higher than in the neutron. Thus the naive perturbation theory
clearly
disagrees with the experimental result 
\cite{Antonelli:1998fv}. 

\begin{figure*}[thb]
\centering
\includegraphics*[width=120mm]{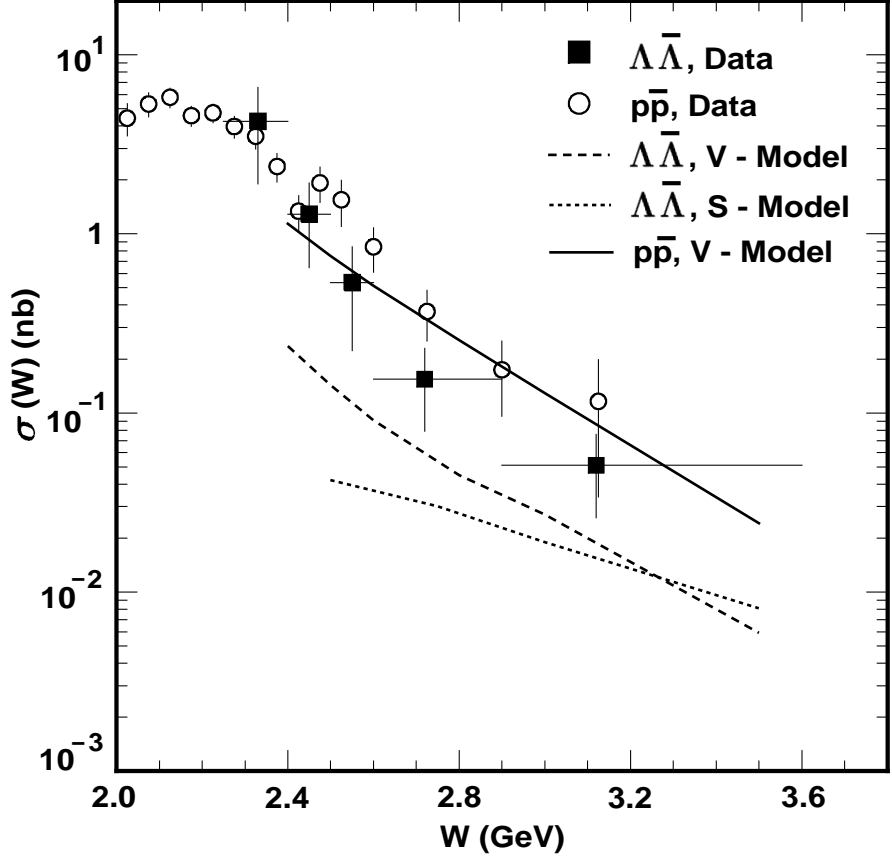}
\caption{
CLEO data 
\cite{Anderson:1997ak}
for
$\sigma_{\gamma\gamma \,\rightarrow\, \Lambda\overline{\Lambda}}(W)$,
$\sigma_{\gamma\gamma \,\rightarrow\, p\overline{p}}(W)$
for $|\cos{\theta^*}| < 0.6$.
Vertical error-bars include
systematic uncertainties.
Horizontal markings indicate bin width.
S-model: scalar quark-diquark model;
V-model: vector quark-diquark model.
}
\label{cleo-gg-fig}
\end{figure*}

\subsection{Where do we go from here?}
As the first step, one has to realize that we are dealing with 
a highly nonperturbative process. Even though the 
\hbox{form} \hbox{factors} are measured at a 
momentum transfer which is much higher than $\Lambda_{QCD}$, \
\hbox{$q^2\gsim 4M_N^2 \sim 4$~GeV$^2$},
we are very far from the perturbative regime. The reason is that all
the available energy is very quickly divided among the quarks and
antiquarks in the $\bar N N$ system. Since the total available 
energy is very close to the rest mass of the  $\bar N N$ system,
none of the quarks has any ``spare" momentum.

There have been several attempts to explain the FENICE data by various
theoretical proposals utilizing specific nonperturbative mechanisms
(for a recent report on some of this work see
Ref.~\cite{Hammer:2001rh}),
but to the best of my knowledge, so far there is 
no satisfactory explanation in terms of conventional mechanisms or their
straightforward extensions. 

The lack of such a conventional theoretical explanation is part of the 
motivation for the proposed new asymmetrical $e^-e^+$ high-statistics
collider at SLAC for the regime $1.4 < \sqrt{s} < 2.5$ GeV \cite{LOI}.
 This machine
will yield high-precision data on baryon production in 
$e^-e^+$ annihilation at threshold, providing a check on the FENICE data and
an accurate benchmark for testing possible theoretical explanations.

In this context
it is amusing to note the perturbation theory predictions for 
the $\Delta$ baryon resonance multiplet production in $e^+e^-$ near
threshold.
Since the $\Delta$ has a
totally symmetric wave function, the corresponding amplitudes 
are equal, $a^\Delta_u=a^\Delta_d\equiv a^\Delta$, 
for all 4 members of the multiplet, 
$\left\{\Delta^{++}, \Delta^{+}, \Delta^{0}, \Delta^{-}\right\}$.
Thus perturbation theory makes a striking prediction 
for the neutral member of the multiplet,
\begin{equation}
\sigma_{\!\scriptscriptstyle PT}
(e^+e^- \,\rightarrow\,  \bar\Delta^{0} \Delta^{0})
\propto
\left\vert
\textstyle
a^\Delta\left({2\over3}-{1\over3}-{1\over3} \right)
\right\vert^2=0\,.
\label{eeDelta}
\end{equation}
More generally, the relative
yields predicted by perturbation theory are
\begin{equation}
\Delta^{++}:\Delta^{+}:\Delta^{0}:\Delta^{-}
\quad = \quad
4:1:0:1\,.
\label{DyieldsPT}
\end{equation}

\section{{$\gamma\gamma\rightarrow \bar N N$}:  A RELATED PUZZLE ?}
The FENICE puzzle is
reinforced by the CLEO data on baryon-antibaryon production 
in photon-photon collisions \cite{Artuso:1994xk}, 
as shown in
Fig.~\ref{cleo-gg-fig}
(see also \cite{L3} and \cite{Hamasaki:1997cy} for related experimental
work).

\begin{figure}[thb]
\centering
\includegraphics*[width=50mm]{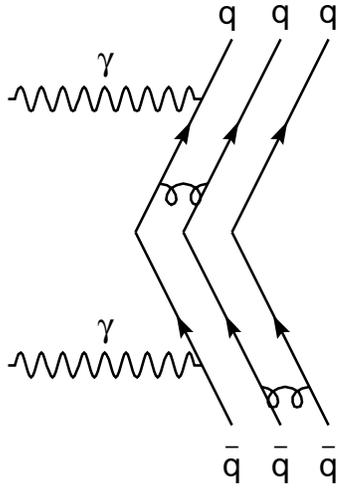}
\caption{Feynman diagram corresponding to the naive perturbative
description for
$\gamma\gamma \,\rightarrow \,\bar N N$.}
\label{gg-fig}
\end{figure}

CLEO has compared the $\gamma\gamma$
cross-sections for $\Lambda\overline{\Lambda}$
and $\bar p p$ production and they find that close to threshold
\hbox{$\sigma(\gamma\gamma \,\rightarrow\, p\overline{p})\approx
 \sigma(\gamma\gamma \,\rightarrow\, \Lambda\overline{\Lambda})$}.
This is quite similar to the FENICE puzzle for the 
$\bar p p/\bar n n$
ratio. The naive perturbative description of the 
baryon-production in the photon-photon reaction is given by the 
Feynman diagram in Fig.~\ref{gg-fig}. Since there are two photons here,
instead of one in Fig.~\ref{ee-fig}, for each flavor of the primary
$\bar q q $ pair the corresponding 
amplitude scales like the quark charge squared, to be compared
with linear dependence of the amplitudes
on the quark charge in the $e^+e^-$ case.
Thus one would naively expect 
the ratio
$\sigma(\bar p p)/\sigma(\overline{\Lambda}\Lambda)$
to be even larger than the corresponding perturbative prediction
for $\sigma(\bar p p)/\sigma(\overline{n}n)$ in $e^+e^-$.

The disagreement between the naive theoretical 
prediction and experiment is striking again.

Just like in the FENICE case, despite several \hbox{attempts},
there is no satisfactory theoretical
explanation for this CLEO data (for example, see
\cite{Berger:1999ri} for a recent theoretical analysis 
of $\gamma\gamma \,\rightarrow\,\hbox{baryons}$
in terms of di-quarks).

It would be highly interesting to see the data for 
$\gamma\gamma \,\rightarrow\, \overline{n} n$ close to threshold,
but such analysis has not yet been done due to some 
technical difficulties \cite{cleo-neutrons}. As will be clear from
the following discussion, close to threshold we expect the
$\gamma\gamma \,\rightarrow\, \overline{n} n$
cross-section be the same as 
$\gamma\gamma \,\rightarrow\, \overline{p} p$.
We urge our experimental colleagues to carry out such an analysis.

The perturbative ratios are even more 
dramatic for the rates for $\gamma \gamma \to \bar\Delta \Delta$,
which are proportional to the fourth power of the $\Delta$ electric charge,
if one has symmetric wavefunctions. Thus for
\ $\sigma_{\!\scriptscriptstyle PT}(\gamma \gamma \to \bar\Delta \Delta)$ \
near threshold,
the analogue of eq.~\ref{DyieldsPT} is
\begin{equation}
\Delta^{++}:\Delta^{+}:\Delta^{0}:\Delta^{-}
\quad = \quad
16:1:0:1\,.
\label{ggDyieldsPT}
\end{equation}
As we shall discuss in more detail in the following section, 
the mechanism we propose for this type of reactions predicts a completely
different result.

\section{THE PROPOSED RESOLUTION}
\begin{figure}[htb]
\centering
\includegraphics*[width=80mm]{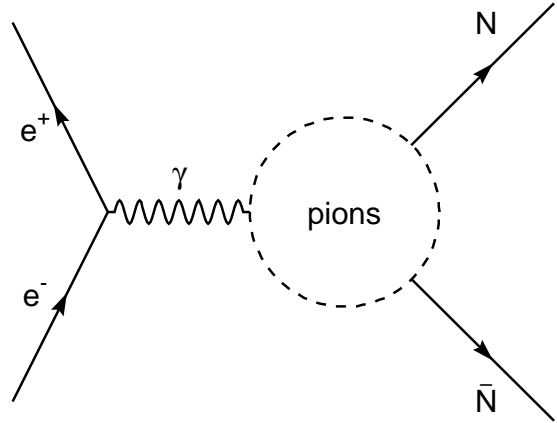}
\caption{$\bar N N\,\rightarrow\,e^+e^-$ as a two-stage process, where
$e^+e^-$ annihilate into a timelike photon which first couples to an
intermediate pion state, which then produces the $\bar N N$ pair.}
\label{pions-fig}
\end{figure}

In view of these FENICE and CLEO
 puzzles, we have proposed \cite{inprog}
a novel mechanism which might explain the data.
The essential new point in the proposed \hbox{mechanism} is that 
it is a two-stage process, with a coherent state of pions serving as an
intermediary between $e^+e^-$ and the baryon-antibaryon system,
as shown schematically in Fig.~\ref{pions-fig}.

\subsection{$\bar N N$ annihilation into pions}
We propose to use the Skyrme model as a concrete computational framework
for a quantitative description of the baryon-antibaryon dynamics.
In the Skyrme model 
\cite{Skyrme:1961vq},\cite{Skyrme:1962vh}
baryons appear as solitons in a purely bosonic chiral
Lagrangian. 
The model is formally justified as a low-energy approximation to
large-$N_c$ QCD
\cite{Witten:1979kh}, \cite{Adkins:1983ya}.
It is known to provide a good description of many
low-energy properties of baryons (see
\cite{Zahed:1986qz} and \cite{Schechter:1999hg} for a review).

It turns out that it is also possible to obtain a fairly accurate description
of low-energy baryon-antibaryon annihilation in terms of
Skyrmion-antiskyrmion annihilation 
\cite{Verbaarschot:1987rj}-\cite{amado}.

\begin{figure}[htb]
\centering
\includegraphics*[width=80mm]{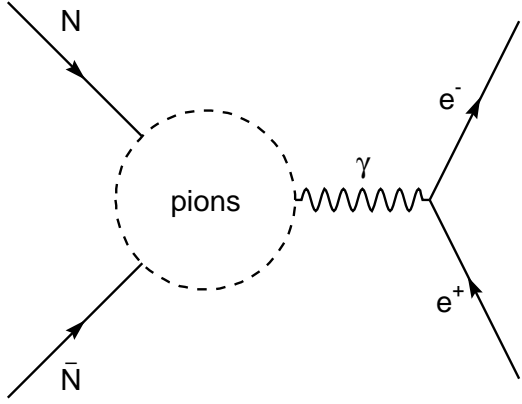}
\caption{Time-reversed process of Fig.~\ref{pions-fig}, i.e.
$\bar N N  \,\rightarrow\,e^+e^-$ as
a two-stage process, where the nucleons first annihilate into 
pions, which then couple to a a timelike photon to produce the 
$e^+e^-$ pair.}
\label{pions1-fig}
\end{figure}

Instead of asking how a \ $\bar p p$ \ or \ $\bar n n$ 
\ configuration is formed
by a virtual timelike photon, or by two photons, it is 
conceptually easier to
consider the reverse processes, i.e. 
\hbox{$\bar p p \,\rightarrow\, \gamma \,\rightarrow\, e^+e^-$}
\
or \ \
\hbox{$\bar n n \,\rightarrow \,\gamma \,\rightarrow\, e^+e^-$},
as shown schematically in Fig.~\ref{pions1-fig},
and the analogous processes for two photons.

It is interesting to note 
that it has been argued \ \cite{Drukier:1982fq} 
that in perturbation theory 
the production of extended \hbox{objects}, such as
a soliton-antisoliton configuration,
by pointlike particles, such as in $e^+e^-$ annihilation,
is suppressed by a large exponential factor 
$\,\sim\exp(-4/\alpha_s)$.
Thus the fact that the reaction $e^+e^-\,\rightarrow\,\bar N N$
has been measured with a small, but finite cross-section
is yet another indication of the nonperturbative nature of the
process.

Following the pioneering numerical work of  
Refs.~\cite{Verbaarschot:1987rj} and \cite{Sommermann:1992yh},
we now have
the following picture of the $N\bar N$ annihilation at rest as a
Skyrmion anti-Skyrmion annihilation:  
just after the Skyrmion and anti-Skyrmion touch, 
they ``unravel" each other, and a
classical pion wave emerges as a coherent burst and
takes away energy and baryon number as quickly as causality permits.

This observation led the authors of Ref.~\cite{Halasz:2001ye}
to suggest the
following simplified version of $N\bar N$ annihilation at rest. After a
very fast annihilation a spherically symmetric "blob" of pionic matter
of size $\sim 1$~Fm, baryon number zero and the total energy twice the
nucleon rest mass is formed. The further evolution of the system and
the branching rates of various channels are completely determined by
the parameters of this "blob".

For a very crude toy model of what is going on, let's assume that
\ $\bar p p$  \
annihilate into two pions which then go to \ $e^+e^-$ \ via a timelike
photon,

\begin{equation}
\bar p p \,\rightarrow\, \pi^+ \pi^- \,\rightarrow
\,\gamma \,\rightarrow \,e^+e^-
\label{pp-2pi-ee}
\end{equation}
Clearly, the real process involves an intermediate state with a 
a much larger number of pions on
the average, but two pions are sufficient to understand why within this
physical picture we expect the
$\bar n n  \,\rightarrow \,e^+ e^-$ rate
to be the same as the
$\bar p p  \,\rightarrow \,e^+ e^-$ rate
rate. The basic
argument is that since we have a two-stage process, the crucial 
issue is the rate
$\bar p p \,\rightarrow\, \pi^+ \pi^-$
vs. the rate
$\bar n n \,\rightarrow \,\pi^+ \pi^-$.
 Since this is a purely strong interaction process, we expect the
two rates to be equal. The next step is 
$\pi^+ \pi^- \,\rightarrow\,  \gamma\, \rightarrow \,e^+e^-$
which of does not care whether the pions were produced by
$\bar p p$ or $\bar n n$ annihilation.
Clearly if 
\ $\bar p p \,\rightarrow\, e^+ e^-$
\ has the same cross-section as 
$\bar n n \,\rightarrow\, e^+ e^-$,
the same will apply to the reverse processes.

It is tempting to  assume that 
a similar argument can be made for 
$\gamma\gamma \,\rightarrow\, \bar\Lambda \Lambda$ \ vs. \
$\gamma\gamma \,\rightarrow \,\bar p p$, \
although one expects the corresponding analysis to be more difficult,
as one will have $K^+K^-$ in the intermediate state.

We should stress that the two pion intermediate state is used here 
only as an illustration. In practice the two pion channel is very small
and most difficult to treat within the approach of
Ref.~\cite{Halasz:2001ye},
since the semiclassical approximation is best suited for coherent states.
A similar comment applies to a possible calculation of 
$\bar N N$ annihilation into two photons \cite{amado}.

In addition to the $e^+e^- \,\rightarrow\, \bar N N$, 
one can also carry out an analogous analysis for
$e^+e^- \,\rightarrow\, \bar \Delta \Delta$.
The relative yields of $\Delta^{++},\Delta^{+},\Delta^{0}$ and $\Delta^{-}$
will be determined by the relevant 
Clebsch-Gordan coefficients \cite{Groom:2000in}
and by the  corresponding reduced matrix elements for isospin 1 and 0.
We do not know the precise values of these reduced matrix elements,
but even without this information, from the Clebsch-Gordan
decomposition we expect
\begin{eqnarray}
BR(\Delta^{++})=BR(\Delta^{-})
\phantom{aaaaaaaaaa}
\nonumber\\
\hbox{and} \phantom{aaaaaaaaaaaaaaaaaaaaaaaaaaaaaaaaaaaaa}
\label{BRratios}
\\
BR(\,\Delta^{+})=BR(\Delta^{0})
\phantom{aaaaaaaaaai}
\nonumber
\end{eqnarray}
as opposed to 
the perturbative prediction (\ref{DyieldsPT}).

Eqs.~(\ref{BRratios}) hold also for 
$\gamma\gamma \rightarrow \bar\Delta \Delta$, since 
$\vert {I_3}^{\Delta^{++}}\vert =
 \vert {I_3}^{\Delta^{-}}\vert$, etc., to be contrasted with
the perturbative prediction (\ref{ggDyieldsPT}).
It would be very interesting to put this
to an experimental test!

\subsection{Time scales of strong vs. EM interactions}
It is not enough to propose a mechanism which can explain the equality of
the observed $\bar p p$ and $\bar n n$ rates. We also have to explain why 
the proposed mechanism dominates over the standard one. After all,
the reaction can in principle still proceed via 
the usual naive perturbative mechanism, where quarks couple directly to the 
virtual photon, like in Fig.~\ref{ee-fig}, and where the $\sigma(\bar p
p)/\sigma(\bar n n) = 3/2.$
Thus a crucial question is why
\ $\bar p p \,\rightarrow\, e^+ e^-$
\ or \ 
\ $\bar p p \,\rightarrow\, e^+ e^-$ \
proceed via intermediate hadronic states, rather
than via direct EM annihilation.  In order to answer this question,
it is helpful to consider the relevant time scales. 

Consider \ $\bar p p$ on top of each other at rest. The QCD annihilation 
occurs at a typical time scale of strong interactions, i.e.
$\sim10^{-24}$ sec. This is much shorter than a typical time scale for EM
interactions, so that the ``direct" QED process, where \ $\bar q q$ \
in the \ $\bar p p$
annihilate into a virtual photon simply has no chance of occurring:
QED here is a ``Johnny come lately" who cannot compete with the QCD rate.

QED enters only at the second stage, where the mesonic ``soup" has a (small)
chance of going into a virtual photon. But here we are concerned with the
{\em relative} rate of $\bar p p$ \ vs. \ $\bar n n$, so the overall
smallness of the QED process $\hbox{pions} \,\rightarrow e^+e^-$ is not a
priori a problem.

\begin{figure*}[t]
\centering
\includegraphics*[width=120mm]{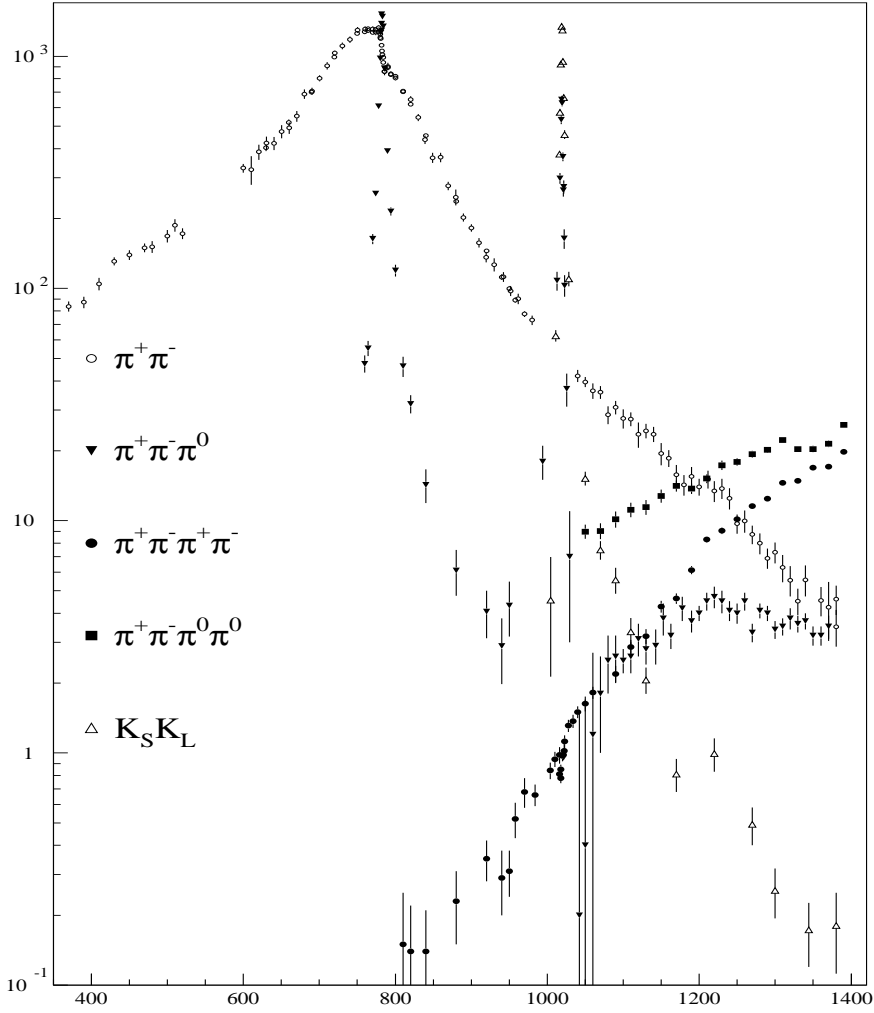}
\caption{$e^+e^- \,\rightarrow\, \hbox{hadrons}$ for $E_{CM} < 1400$ MeV;
\ recent data from Novosibirsk VEPP-2M collider \cite{bondar}.}
\label{vepp-fig} 
\end{figure*}

\subsection{A more realistic intermediate state}
The two-pion intermediate state is of course only a toy model which is
helpful in understanding the qualitative features of the reaction. In order
to obtain a more more realistic description, we need to put in a more
realistic intermediate state. The first improvement would be to include 
intermediate states with $n$ pions, where $n$ goes over all allowed
values,
\begin{equation}
\sigma(\bar p p\rightarrow e^+e^-) \sim 
\left\vert \sum_n \langle \bar p p \vert n \pi \rangle 
\langle n \pi\vert e^+e^- \rangle 
\right\vert^2
\label{n-pions}
\end{equation}

In principle one could compute the $\bar p p \rightarrow n \pi$ rates
using the methods of Ref.~\cite{Halasz:2001ye}. 
It is not clear how to obtain the relative phases for different values of
$n$, but if one of the intermediate states dominates, this problem
will not be of practical significance.

A more sophisticated treatment will involve summing over all allowed 
intermediate states, not just the $n$-$\pi$ \hbox{states}. 
In principle this could
be done by combining the low-energy $e^+e^- \rightarrow\hbox{hadrons}$
data with the corresponding data for 
$\bar N N $ annihilation at rest. One would then sum them channel by
channel. 
There are very precise data 
from LEAR for $\bar N N $ annihilation a rest and there are also good data
for $e^+e^- \rightarrow\hbox{hadrons}$. Recently the VEPP-2M collider at
Novosibirsk provided highly accurate $e^+e^-$ data for $E_CM < 1.4$~GeV 
\cite{bondar},
with excellent finite state resolution, as
shown in Fig.~\ref{vepp-fig}.

This energy range is below what we need, but it shows the expected
richness of the data. If similar data can be obtained for $E_{CM}\gsim2$
GeV, they could be combined with the LEAR data to provide an estimate
of the $\bar p p\rightarrow e^+e^-$ rate through our mechanism.

Again, one remaining difficulty is the issue of relative phases, but 
as already mentioned, if 
the intermediate state is dominated but one particular channel, the phase
issue will not be of practical significance.

\subsection{A caveat}
Clearly, what is presented here is
merely a sketch of the proposed calculation, and in order to convince oneself
that it correctly describes the physics, one should actually 
put in the rates for the relevant
intermediate processes, in order to provide a theoretical estimate which can be
compared with the measured rate for \
$e^+e^- \,\rightarrow\, \bar p p $ \ or \
$e^+e^- \,\rightarrow\, \bar n n $.

\section{ACKNOWLEDGEMENTS}
The work described in this talk was carried out in collaboration with Stan
Brodsky. We are grateful to Rinaldo Baldini for many discussions of the
FENICE data.
We also wish to thank 
Ralph Amado, Alexander Bondar and Hans P. Paar
for communicating and \hbox{discussing} their results with us.
This   research was supported in part
by a grant from the United States-Israel
Binational Science Foundation (BSF), Jerusalem, Israel,
and by the Basic Research Foundation administered by the
Israel Academy of Sciences and Humanities.

\end{document}